\documentstyle[12pt]{article}

\begin{document}

\title{Stabilization of the Electroweak String by the Bosonic Zero Mode}
\author{V.B. Svetovoy \thanks{%
e-mail: vs@postoff.ics.ac.ru}  \\ 
{\small Department of Physics, Yaroslavl State University,} \\ {\small %
Sovetskaya 14, Yaroslavl 150000, Russia}}
\date{}
\maketitle

\begin{abstract}
In the {\it minimal }standard model we discuss stability of the $Z$-string
configuration, which includes the zero mode connected with the broken gauge
symmetry. The zero mode induces charge and current on the string and gives
backreaction to the string profile changing the region of dynamical
stability. For $\sin ^2\theta _W=0.23$ it is found that the string is
stabilized for the Higgs mass $M_H<132\ GeV$ in some finite range of the
time-like currents.
\end{abstract}


\vspace{0.5cm}

The Nielsen-Olesen vortex solution of the abelian Higgs model \cite{NO} can
be embedded into the $SU(2)_L\times U(1)_Y$ electroweak theory \cite{V92}
and this solution is known as the electroweak or $Z$-string. It could
trigger baryogenesis at the electroweak phase transition \cite{BD93} and
manifest itself in accelerator experiments. However, there are no
topological arguments for stability of such a string. The dynamical
stability is reached outside of the physical region of the parameters \cite
{JPV93,P93,GH95}. Different attempts have been undertaken to save the $Z$%
-strings. It was shown that bound states on the string may improve its
stability \cite{VW92}. Fermion zero modes do not provide the backreaction to
the string forming fields \cite{EP94} leaving the string unstable.
Stabilization is possible in different extensions of the standard model
where even topologically stable strings can be formed \cite{DS93}, but here
we will consider the minimal model.

Any string solution inevitably breaks some internal and space-time
symmetries generating massless excitations of the string (''zero modes'')
connected with the broken transformations. In some cases these excitations
are very important for the string dynamics. A well known example is the
string superconductivity in the Witten model \cite{W85} caused by the
bosonic zero mode originated from the broken electromagnetic gauge symmetry
inside the string core. Recently it has been shown that the Nielsen-Olesen
string also has the zero mode of this kind \cite{S97} though the gauge
symmetry is unbroken in the string center. It has been argued that the
string fields interpolating between false and true vacua definitely break
the gauge symmetry in the core inducing the zero mode with specific
properties. This mode provides the backreaction to the string forming
fields, induces charge and current on the string, and changes the excitation
spectrum. As a result it will also change the region of dynamical stability
for the $Z$-string.

In this paper the stability region for this current-carrying $Z$-string is
analyzed. The string solution with a ground state zero mode is described
first and its influence on the string profile is calculated numerically.
Then the excitation spectrum is investigated and the range of stability is
mapped on the Higgs mass - current plane for the physical value of the
Weinberg angle.

The $Z$-string is the Nielsen-Olesen string embedded into the electroweak
theory by such a way that the upper component of the Higgs doublet $\Phi
_u=0 $ and the gauge bosons $W_\mu ^{\pm }=A_\mu =0$. The $Z$-boson and the
down component of the doublet $\Phi _d$ are described by the lagrangian of
the abelian Higgs model

\begin{equation}
\label{L1}{\cal L}=-\frac 14Z_{\mu \nu }Z^{\mu \nu }+(D_\mu \Phi
_d)^{*}(D_\mu \Phi _d)-\lambda \left( \left| \Phi _d\right| ^2-\eta
^2/2\right) ^2,
\end{equation}

\noindent where

$$
Z_{\mu \nu }=\partial _\mu Z_\nu -\partial _\nu Z_\mu ,\quad D_\mu \Phi
_d=\left( \partial _\mu +iqZ_\mu \right) \Phi _d.
$$

\noindent The charge $q$ is expressed via coupling constants $g$ and $%
g^{\prime }$ of the groups $SU(2)_L$ and $U(1)_Y$, respectively, as $q=\sqrt{%
g^2+g^{\prime 2}}/2$ and masses of the $Z$ and Higgs bosons are

\begin{equation}
\label{mas}M_Z=q\eta ,\quad M_H=\sqrt{2\lambda }\eta .
\end{equation}

\noindent Let us consider the following ansatz for a unit winding string
along the $z$-axis \cite{S97}

\begin{equation}
\label{str}Z_i=\partial _i\varphi _0(x)f_0(r),\quad Z_\alpha =\frac
1{qr^2}\epsilon _{\alpha \beta }y_\beta Z(r),{\cal \quad }\Phi =\frac \eta {%
\sqrt{2}}f(r)e^{i\vartheta },
\end{equation}

\noindent where the notations $x^i=(t,z)$ for $i=0,1$ and $y^\alpha =(r\cos
\vartheta ,r\sin \vartheta )$ for $\alpha =1,2$ were introduced for
longitudinal and transverse coordinates, respectively. If the function $%
\varphi _0(x)=0$, one gets the standard vortex solution \cite{NO}. $%
Z_i=\partial _i\varphi _0(x)f_0(r)$ will describe a zero mode of this vortex
connected with the broken gauge symmetry in the string core if the function $%
f_0(r)$ obeys the equation

\begin{equation}
\label{zmeq}f_0^{\prime \prime }+\frac 1rf_0^{\prime }-M_Z^2f^2f_0=0.
\end{equation}

\noindent One can check that in this case the equations of motion give $%
\partial _i\partial ^i\varphi _0(x)=0$ and, therefore, $\varphi _0(x)$ is a
2-dimensional massless field. The problem is that solutions of Eq.(\ref{zmeq}%
) are always singular at infinity or at the origin. If one supposes that $%
Z_i $ obeys usual boundary condition $Z_i\rightarrow 0$ at $r\rightarrow
\infty $ and $Z_i$ is restricted at $r\rightarrow 0$, then one has to
exclude the mode as having infinite energy. The boundary conditions for this
zero mode of the Nielsen-Olesen string have been analyzed in Ref.\cite{S97},
where it was found that $Z_i$ is restricted by the condition

$$
\int d^2xd^2y\partial _\alpha \left( Z_i\partial _\alpha Z^i\right) =
$$

\begin{equation}
\label{bc2}2\pi \int d^2x\partial _i\left( \varphi _0\partial ^i\varphi
_0\right) \int dr\frac d{dr}\left( rf_0f_0^{\prime }\right) =0.
\end{equation}

\noindent Now we see that there is another way to meet (\ref{bc2}): one can
restrict the behavior of the field $\varphi _0(x)$ at the space-time $(t,z)$
boundaries in usual way instead of bound the $r$-behavior of $f_0(r)$. It
was shown that $\varphi _0(x)$ will be normalized as a 2-dimensional field
if $f_0(r)$ obeys the relaxed boundary conditions allowing logarithmic
singularity at the origin:

\begin{equation}
\label{bc3}f_0(r)f_0^{\prime }(r)=-\frac 1{2\pi r},\ r=r_0\rightarrow
0;\quad f_0(r)\rightarrow 0,\ r\rightarrow \infty .
\end{equation}

\noindent Here the mathematical cutoff parameter $r_0\rightarrow 0$ has been
introduced.

The mode $\varphi _0(x)$ carries a finite energy

\begin{equation}
\label{ezm}E_{zm}=\frac 12\int\limits_0^Ldx_1\left[ \left( \partial
_0\varphi _0\right) ^2+\left( \partial _1\varphi _0\right) ^2\right] ,
\end{equation}

\noindent where $L$ is the string length, and propagating along the string
with the speed of light. $E_{zm}$ does not increase with $L$ although the
string solution has finite energy per unit length and, of course,
excitations with finite energy per unit length should be permitted. This
means that the functions $\varphi _0(x)$ with linear singularities at the
space-time boundaries are also allowed. The mode $\varphi _0(x)$ which is
linear in time and position obeys the equation of motion and carries the
energy proportional to the string length $L$. It cannot be expanded in
harmonic oscillators since it obeys different boundary conditions. For this
reason we have to attribute such a mode to the background solution rather
than the string perturbations. In this sense the mode is quite similar to
the explicit collective coordinate \cite{R82}. For the background value of
the zero mode one takes

\begin{equation}
\label{phi}\varphi _0(x)=b_ix^i=b_0t-b_1z,
\end{equation}

\noindent where $b_i$ is a constant 2-vector. In this case the string is
described by a time and position independent field configuration and its
energy is

\begin{equation}
\label{eng}E=E_{core}+E_{zm},
\end{equation}

$$
E_{core}=\pi \eta ^2L\int\limits_0^\infty d\rho \rho \left\{ f^{\prime 2}+%
\frac{Z^{\prime 2}}{\rho ^2}+\frac{\left( 1-Z\right) ^2}{\rho ^2}f^2+\frac
14\beta \left( f^2-1\right) ^2\right\} ,
$$

$$
E_{zm}=\frac{\eta ^2}%
2L(a_0^2+a_1^2),
$$

\noindent where the dimensionless variables were introduced

\begin{equation}
\label{resc}\rho =M_Zr,\quad \beta =\frac{M_H^2}{M_Z^2},\quad a_i=\frac{b_i}%
\eta ,
\end{equation}

\noindent and primes denote differentiation with respect to $\rho $.

A string solution similar to (\ref{str}) with $\varphi _0(x)$ as in (\ref
{phi}) has been proposed \cite{KLY97} for a specially constructed
non-abelian string with the only important difference that the solution was
singular at infinity instead of the origin as in our case. There is also
some resemblance of (\ref{str}) and the dion solution of the $O(3)$ model
\cite{R82} which, of course, comes from the fact that both solutions are
connected with the zero mode of the same origin.

Solving Eq.(\ref{zmeq}) with the conditions (\ref{bc3}) one finds that for
small $\rho $ the $Z_i$ components of the vector potential are

\begin{equation}
\label{pot}Z_i=\eta a_i\frac{\ln (\overline{\rho }/\rho )}{\sqrt{2\pi \ln (%
\overline{\rho }/\rho _0)}},\quad \rho \ll 1,
\end{equation}

\noindent where $\overline{\rho }\sim 1$ is a constant. The first impression
is that there are sources in the string center generating this potential.
However, it is false because the current density induced by the mode

\begin{equation}
\label{cd}j_i=-M_Z^2\eta a_if_0(\rho )f^2(\rho )
\end{equation}

\noindent disappears as $\rho ^2\ln \rho $ for small $\rho $. A real density
has a maximum at $\rho \sim 1$ and goes to zero at both boundaries.
Integrating (\ref{cd}) over the string cross section with the help of (\ref
{zmeq}) and (\ref{bc3}) one finds for the charge per unit length $J_0$ and
the current $J_1$ along the string

\begin{equation}
\label{cur}J_i=-\frac{a_i\eta }{f_0(\rho _0)}=-a_i\eta \sqrt{\frac{2\pi }{%
\ln (\overline{\rho }/\rho _0)}}.
\end{equation}

\noindent In what follows we will often save the term ''current'' for both
components of $J_i$.

In the mathematical limit $\rho _0\rightarrow 0$ the current goes to zero
and the physical effect of the zero mode disappears. In physical reality we
cannot believe in our equations for arbitrary small distances. A natural
candidate for the cutoff parameter $r_0$ is the Plank length $M_P^{-1}$. It
cannot be any scale of symmetry breaking in Grand Unification Theory since
the field equations are still valid. However, if the particles in the
standard model are composite, the compositness scale will play the role of $%
r_0$. Further we will suppose that $\rho _0=M_Z/M_P$. The zero mode is
concentrated at small distances $\sim 1/M_P$ but survives at $r\sim 1/M_Z$
because it decreases logarithmically.

Let us investigate now the backreaction of the zero mode to the string
profile. Equations of motion following from (\ref{L1}) give for the
functions in our string ansatz

$$
f^{\prime \prime }+\frac 1\rho f^{\prime }-\frac{(1-Z)^2}{\rho ^2}f+\gamma
f_0^2f-\frac 12\beta \left( f^2-1\right) f=0,
$$

$$
Z^{\prime \prime }-\frac 1\rho Z^{\prime }+\left( 1-Z\right) f^2=0,
$$

\begin{equation}
\label{prof}f_0^{\prime \prime }+\frac 1\rho f_0^{\prime }-f^2f_0=0,
\end{equation}

\noindent where $\gamma =a_0^2-a_1^2$. These equations have to be solved
with the boundary conditions

$$
\rho =\rho _0:\qquad f=0,\quad Z=0,\quad f_0f_0^{\prime }=-\frac 1{2\pi \rho
_0};
$$

\begin{equation}
\label{bc4}\rho \rightarrow \infty :\qquad f\rightarrow 1,\quad Z\rightarrow
1,\quad f_0\rightarrow 0,
\end{equation}

\noindent There is no way to analyze the effect analytically, so one needs
numerical investigation. The problem is that $\rho _0\sim 10^{-17}$ is too
small for numerical study. The way out is to shift the boundary condition to
a point $\rho =\rho _1$, where $\rho _1$ is not too small. This can be done
since for small $\rho $ one can solve Eq.(\ref{prof}) analytically. Doing
this up to the terms $\sim \rho ^2$ one finds nonlinear conditions at $\rho
=\rho _1$:

$$
\rho _1f^{\prime }(\rho _1)s_1=f(\rho _1)s_2,\quad \rho _1Z^{\prime }(\rho
_1)=2Z(\rho _1),
$$

\begin{equation}
\label{bc5}f_0(\rho _1)=p\sqrt{\frac{\ln \left( \rho _1/\rho _0\right) }{%
2\pi \left( 1-p\right) }},\quad p=-2\pi \rho _1f_0(\rho _1)f_0^{\prime
}(\rho _1),
\end{equation}

\noindent where

$$
s_1=1-\frac{\rho _1^2\overline{J}^2}8\left[ \frac 78+\frac \beta {2\overline{%
J}^2}+\frac 32\ln \left( \frac{\bar \rho }{\rho _1}\right) +\ln ^2\left(
\frac{\bar \rho }{\rho _1}\right) \right] ,
$$

$$
s_2=1-\frac{\rho _1^2\overline{J}^2}8\left[ \frac 98+\frac{3\beta }{2%
\overline{J}^2}+\frac 52\ln \left( \frac{\bar \rho }{\rho _1}\right) +3\ln
^2\left( \frac{\bar \rho }{\rho _1}\right) \right] ,
$$

$$
\overline{J}^2=\frac{J_0^2-J_1^2}{\left( 2\pi \eta \right) ^2}=\frac \gamma
{2\pi \ln \left( \bar \rho /\rho _0\right) },
$$

\begin{equation}
\label{deff}\ln \left( \frac{\overline{\rho }}{\rho _1}\right) =\pi
f_0^2(\rho _1)\left[ 1+\sqrt{1+\frac{2\ln \left( \rho _1/\rho _0\right) }{%
\pi f_0^2(\rho _1)}}\right] .
\end{equation}

\noindent The last two expressions define the relation between the current
and the parameter $\gamma $ which is more suitable to use instead of the
current. For practical purposes one can take $\overline{\rho }=2 $ at least
in the range $0.25<\beta<4 $.

The shift of the boundary allows to take the value of $\rho _1$ as large as $%
0.1$. For the calculations it was chosen $\rho _1=0.05$. Even after this
procedure a linear discretization of $\rho $ is not appropriate, so the
variable $\xi =\ln \rho $ was used. Equations (\ref{prof}) were discretized
on a lattice with 32, 64 or 128 points and the resulting system of nonlinear
equations was solved by the Newton method. The results practically
insensitive to the number of points in the range. Comparison of the linear
and logarithmic discretization for the zero current showed a good agreement.
After the solution has been found the independent check has been done with
the Runge-Kutta algorithm. The results for $\beta =1$ and different values
of $\gamma $ are shown in Fig.1. The width of the scalar core is increased
with $\left| \gamma \right| $ for the space-like current ($\gamma <0$) and
the symmetry is restored near the center. For the time-like current ($\gamma
>0$) the Higgs field width is decreased with $\gamma $ increase. In both
cases $Z(\rho )$ and $f_0(\rho )$ are much less sensitive to the value of $%
\gamma $. There is no natural restriction on the current value in contrast
with the case of the Witten string \cite{DS88}.

One considers small excitations around the classical solution (\ref{str}).
Our goal is to find the stability conditions and for this reason one can set
$\delta Z_\mu $ and $\delta \Phi _d$ to be zero because the $U(1)$ string is
topologically stable and the zero mode cannot spoil the conclusion. The
string solution does not break the electromagnetic $U(1)_{em}$ symmetry,
therefore, $\delta A_\mu $ is not important for the stability problem. For
the rest of the excitations one chooses the notations

\begin{equation}
\label{defex}\delta W_\mu ^{+}=V_\mu ^{+},\qquad \delta W_\mu ^{-}=V_\mu
^{-},\qquad \delta \Phi _u=\Psi .
\end{equation}

\noindent In the description of these fields let us follow to the method
developed by Goodband and Hindmarsh \cite{GH95} and choose the background
gauge condition

\begin{equation}
\label{gc}\partial ^\mu V_\mu ^{+}-ig\cos \theta _WZ^\mu V_\mu ^{+}-i\frac g{%
\sqrt{2}}\Phi _d^{*}\Psi =0,
\end{equation}

\noindent In this gauge the fields $V_\mu ^{+}$ and $\Psi $ will obey the
linearized equations of motion

\begin{equation}
\label{Weq}{\cal D}_\mu {\cal D}^\mu V_\nu ^{+}+2ig\cos \theta _WZ_{\mu \nu
}V^{+\mu }-i\sqrt{2}g\left( \bar D_\nu \Phi _d\right) ^{*}\Psi +\frac
12g^2\left| \Phi _d\right| ^2V_\nu ^{+}=0,
\end{equation}

\begin{equation}
\label{psieq}D_\mu D^\mu \Psi -i\sqrt{2}gV^{+\mu }\bar D_\mu \Phi _d+\left[
\frac 12g^2\left| \Phi _d\right| ^2+2\lambda \left( \left| \Phi _d\right|
^2-\eta ^2/2\right) \right] \Psi =0,
\end{equation}

\noindent where

\begin{equation}
\label{def}\bar D_\mu =\partial _\mu +iqZ_\mu ,\ {\cal D}_\mu =\partial _\mu
-2iq\cos ^2\theta _WZ_\mu ,\ D_\mu =\partial _\mu -iq\cos 2\theta _WZ_\mu .
\end{equation}

\noindent $V_\mu ^{-}$ obey the conjugated equations. The gauge condition
allows the residual gauge freedom with a gauge functions $\chi ^{+}$ which
is a solution of the equation

\begin{equation}
\label{ghost}{\cal D}_\mu {\cal D}^\mu \chi ^{+}+\frac 12g^2\left| \Phi
_d\right| ^2\chi ^{+}=0.
\end{equation}

\noindent This means that the gauge fixing term must be accompanied by
Fadeev-Popov ghost terms. Two of 5 fields $V_\mu ^{+}$ and $\Psi $ are
canceled by the ghost and the other 3 correspond to the spin states of the
massive $W$ boson.

Of the total of 5 equations (\ref{Weq}),(\ref{psieq}) only 4 are coupled. To
see it, one introduces the combinations

\begin{equation}
\label{VU}V^{+}=\frac{a^iV_i^{+}}a,\qquad U^{+}=\frac{\epsilon
^{ij}a_jV_i^{+}}a.
\end{equation}

\noindent where $a=\sqrt{\left| a^ia_i\right| }$. Then the field $U^{+}$
decouples and is canceled by the ghost. The other unphysical field is some
linear combination of the remaining 4 fields. Since the string configuration
is time and position independent, the $x$-dependence of the fields can be
sought in the form $e^{-i\omega t+ikz}$. To solve the equations, one expands
the fields in total angular momentum states

$$
\Psi (x,y)=e^{-i\omega t+ikz}\sum_m\psi ^m(\rho )e^{i\left( m+1\right)
\vartheta },
$$

$$
V_u^{+}(x,y)=e^{-i\omega t+ikz}\sum_m-iv_u^m(\rho )e^{i\left( m-1\right)
\vartheta },
$$

$$
V_d^{+}(x,y)=e^{-i\omega t+ikz}\sum_miv_d^m(\rho )e^{i\left( m+1\right)
\vartheta },
$$

\begin{equation}
\label{Ydef}V^{+}(x,y)=e^{-i\omega t+ikz}\sum_mv^m(\rho )e^{im\vartheta },
\end{equation}

\noindent where $V_{u,d}^{+}=\frac 1{\sqrt{2}}\left( V_1^{+}\mp
iV_2^{+}\right) $ are spin up and spin down states. Substituting it into (%
\ref{Weq}) and (\ref{psieq}) one finds the eigenvalue problem

\begin{equation}
\label{Yeq}{\bf Y}_m^{\prime \prime }+\frac 1\rho {\bf Y}_m^{\prime }+{\cal M%
}^2(\omega ,k,m,\rho ){\bf Y}_m=0,
\end{equation}

\noindent where the vector ${\bf Y}_m=\left( \psi ^m,v_u^m,v_d^m,v^m\right) $
and the matrix ${\cal M}^2$ is

\begin{equation}
\label{matr}{\cal M}^2(\omega ,k,m,\rho )=\left(
\begin{array}{cccc}
D_1 & A & B & -C \\
A & D_2 & 0 & -E \\
B & 0 & D_3 & E \\
sign(\gamma )C & sign(\gamma )E & -sign(\gamma )E & D_4
\end{array}
\right) .
\end{equation}

\noindent The matrix elements are defined by the relations

$$
D_1=\left( \omega +a_0c_2f_0\right) ^2-\left( k+a_1c_2f_0\right) ^2-\frac{%
\left( m+1+c_2Z\right) ^2}{\rho ^2}-c_1^2f^2-\frac 12\beta \left(
f^2-1\right) ,
$$

$$
D_2=\left( \omega +2a_0c_1^2f_0\right) ^2-\left( k+2a_1c_1^2f_0\right) ^2-%
\frac{\left( m-1+2c_1^2Z\right) ^2}{\rho ^2}-c_1^2f^2-\frac{4c_1^2Z^{\prime }%
}\rho ,
$$

$$
D_3=\left( \omega +2a_0c_1^2f_0\right) ^2-\left( k+2a_1c_1^2f_0\right) ^2-%
\frac{\left( m+1+2c_1^2Z\right) ^2}{\rho ^2}-c_1^2f^2+\frac{4c_1^2Z^{\prime }%
}\rho ,
$$

$$
D_4=\left( \omega +2a_0c_1^2f_0\right) ^2-\left( k+2a_1c_1^2f_0\right) ^2-%
\frac{\left( m+2c_1^2Z\right) ^2}{\rho ^2}-c_1^2f^2,
$$

$$
A=-\sqrt{2}c_1\left( f^{\prime }-\frac{1-Z}\rho f\right) ,\ B=\sqrt{2}%
c_1\left( f^{\prime }+\frac{1-Z}\rho f\right) ,
$$

\begin{equation}
\label{mel}C=2c_1af_0f,\quad E=2\sqrt{2}c_1^2af_0^{\prime }.
\end{equation}

\noindent Here we denoted $c_1=\cos \theta _W$, $c_2=\cos 2\theta _W$ and
introduced dimensionless $\omega \rightarrow \omega /M_Z$ and $k\rightarrow
k/M_Z$. The functions ${\bf Y}_m(\rho )$ are restricted at the origin and
vanish at infinity. Of course, the condition at $\rho =0$ has to be shifted
to the point $\rho =\rho _1$ by solving (\ref{Yeq}) analytically for small $%
\rho $ as we did for the string profile.

The string with zero current is unstable for $m=-1$ mainly due to
interaction of the string ''magnetic'' field with $v_d$ (the term $%
4c_1^2Z^{\prime }/\rho $ in $D_3$). The zero mode gives rise to the
transverse components of the ''electric'' and ''magnetic'' fields confined
in the string core ($Z_{\alpha i}$ components of the field tensor).
Interaction of the excitations with these fields has non-abelian nature and
plays a crucial role in the stability problem. In ${\cal M}^2$ the elements
responsible for this interaction are ${\cal M}_{24}^2=\pm {\cal M}_{42}^2$
and ${\cal M}_{34}^2=\pm {\cal M}_{43}^2$. They are important because behave
as $1/\rho $ for small $\rho $. For the space-like current ($\gamma <0$)
these elements are symmetric and then the transverse components of the field
strength work against the stability. On the contrary, the time-like current (%
$\gamma >0$) ensures that the matrix elements are antisymmetric what
improves the stability if the current is not too large. That is because the
symmetric elements reduce the minimal eigenvalue of ${\cal M}^2$ but the
antisymmetric ones enlarge it up to the moment it becomes complex.
Therefore, we expect some finite stability region for time-like currents in
the string. This simple picture is spoiled by the components of the zero
mode potential $Z_i$ which are combined with $\omega $ and $k$ in (\ref{mel}%
) and by the first component of ${\bf Y}$ which is not governed by the field
strength at all. Nevertheless, the calculations support this qualitative
description.

Our calculations for the string with zero current agree well with the
results of the previous works \cite{JPV93,GH95}. All the other calculations
were made for the physical value of $\sin ^2\theta _W=0.23$. It was checked
that the pure current ($J_0=0$) makes lower the minimal eigenvalue $\omega
^2 $ of the problem (\ref{Yeq}) which is always negative in the $m=-1$
sector. For the pure charge on the string ($J_1=0$) a finite range of $%
\gamma $ for which the lowest level $\omega $ is real was found for a fixed $%
\beta $ (see Fig.2). The level corresponds to a bound state. In this range
the string is stable. The two lowest levels merge at the end points of the
stability interval and become complex outside of this interval. When $\beta $
increases the range of stability shrinks. For $\beta >2.1$ the string cannot
be stabilized at any current and, therefore, the stable electroweak string
can exist only if the Higgs mass is smaller than $132\ GeV$. The separate
''star'' in Fig.2 shows the result for the left boundary at $\beta =0.02$
which is the smallest value we were able to consider. The range of small $%
\beta $ is important for description of the phase transition. The equations (%
\ref{Yeq}) were also investigated for $m=0$ and $m=1$ where instabilities
are possible in principle but we not found any.

We have considered only the bosonic sector of the standard model. Naculich
\cite{N95} has analyzed the effects of fermions on the stability of the $Z$%
-string and has found that the fermions destabilize the string for all
values of the parameters. The arguments given by Liu and Vachaspati \cite
{LV96} convince, however, that fermionic vacuum instability should lead to a
distortion of the bosonic string but not be responsible for decay. These
arguments are quite common and have to be true for the string configuration
considered in this paper. Therefore, our conclusion is that for $M_H<132\ GeV
$ there is a finite range of time-like currents for which the electroweak
string is stable.

\newpage
\centerline{\large \bf Figure Captions}

\vspace{5mm}

\noindent {\bf Figure 1}. The string profile for $\beta =1$ and different
values of $\gamma $. The functions $f(\rho )$, $Z(\rho )$, and $f_0(\rho )$
are represented as dotted, dashed, and solid lines, respectively.

\vspace{5mm}

\noindent {\bf Figure 2}. Stability region of the string (under the curve).
The ''stars'' are the points of actual calculations. The separate ''star''
corresponds to the smallest value of $\beta =0.02$. The relation of $\gamma $
with the current is defined by Eq.(\ref{deff}).

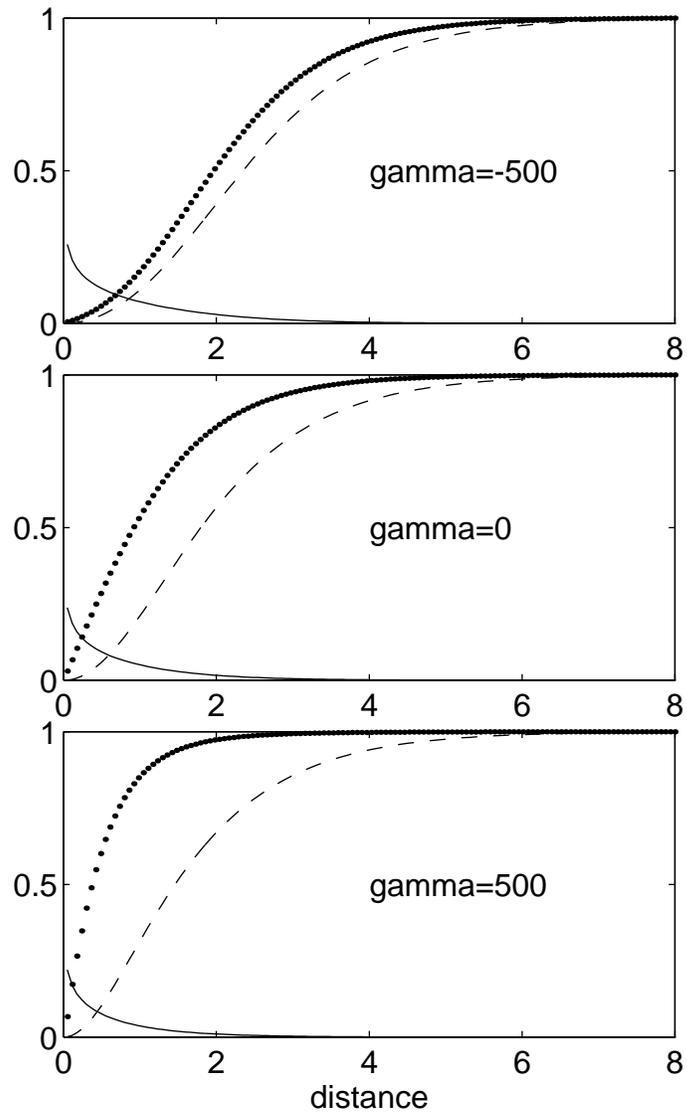
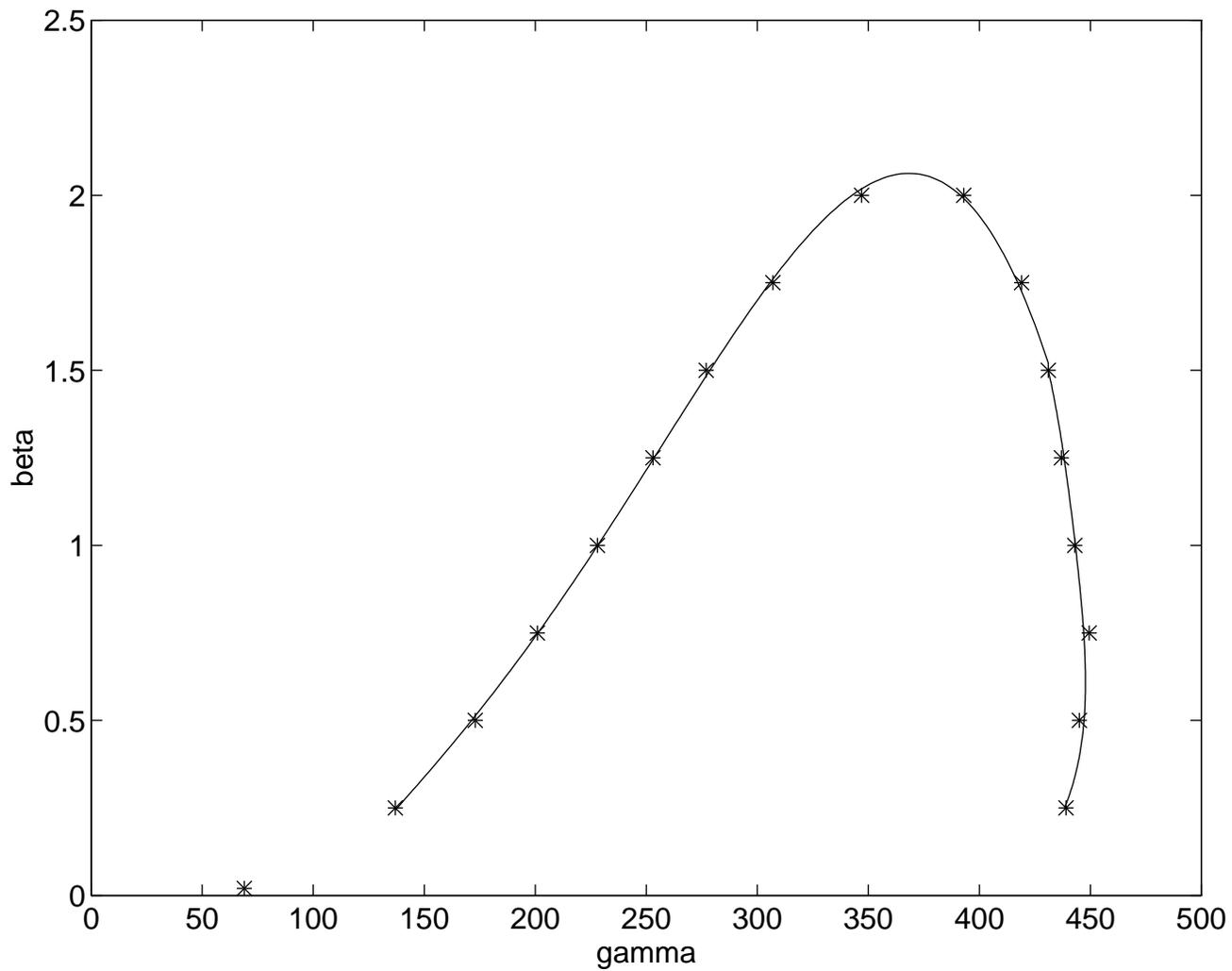
\end{document}